\documentclass[twocolumn]{aa}  

\usepackage{soul}
\usepackage{graphicx}
\usepackage{amsmath}
\usepackage{soul}
\usepackage{hyperref}
\hypersetup{
    colorlinks=true,
    allcolors=blue,
}
\usepackage{txfonts}

\begin{document}

   \title{The Outer spiral arm of the Milky Way using Red Clump stars}
   \subtitle{Tracing the asymmetry across the Disc}

   \author{Namita Uppal
          \inst{1,2}
          \and
          Shashikiran Ganesh\inst{1}
          \and 
          Mathias Schultheis\inst{3}
          }

   \institute{Physical Research Laboratory, Ahmedabad, 380009, Gujrat, India\\
              \email{namita@prl.res.in}
         \and
             Indian Institute of Technology, Gandhinagar, 382355, Gujrat, India\\
             \email{namita.uppal@iitgn.ac.in}
        \and
            Université Côte d’Azur, Observatoire de la Côte d’Azur, CNRS, Laboratoire Lagrange, Bd de l’Observatoire, CS 34229, 06304 Nice cedex 4, France\\
             }

  \abstract
   {}
   {Our aim is to provide an observational view of the old Disc structure of the Milky Way galaxy using the distribution of red clump stars. The spiral arms, warp structure, and other asymmetries present in the Disc are re-visited using a systematic study of red clump star counts over the disc
   of the Galaxy.}
   {We developed a method to systematically extract the red clump stars from 2MASS ($J-K_s, ~J$) colour-magnitude diagram of $1^\circ \times 1^\circ$ bins in $\ell \times b$ covering the range $40^\circ \le \ell \le 320^\circ$ and $-10^\circ \le b \le 10^\circ$. 2MASS data continues to be important since it is able to identify and trace the red clump stars to much farther distances than any optical survey of the Disc. The foreground star contamination in the selected sample is removed by utilising the accurate astrometric data from Gaia EDR3. We analysed the spiral arms and asymmetry in the Galaxy above and below the Galactic plane in the Galactocentric coordinates. }
   {We have generated a face-on-view  (XY-plane) of the Galaxy depicting the density distribution and count ratio above and below the Galactic plane.  The resulting over-density of red clump stars traces the continuous morphology of the Outer arm from the second to the third Galactic quadrant. This is the first study to map the Outer arms across the disc using red clump stars.  Through this study, we are able to trace the Outer arm well into the 3rd Galactic quadrant for the first time. Apart from the spiral structures, we also see a wave-like asymmetry above and below the Galactic plane with respect to longitudes indicating the warp structure. The warp structure is studied systematically by tracing the ratio of red clump stars above and below the Galactic plane.  We provide the first direct observational evidence of the asymmetry in the Outer spiral arms confirming that the spiral arms traced by the older population are also warped, similar to the Disc.
   }
   {}


   \maketitle
%

\section{Introduction}
Studying the Galactic structure is crucial for understanding the formation and evolution  of the Milky Way galaxy. Our location within the Disc makes it challenging to obtain a detailed picture of its structure. Interstellar dust and projection effects, as well as the multifaceted
characteristics of the arms themselves interfere with the determination of spiral arm properties. Mapping 3D structures using star counts is widely used for this purpose \citep{paul}, and the importance of this tool has increased in the past decades with the advent of wide-field surveys. In the last decade, many successful efforts have been dedicated to increase the understanding of the structure, morphology and dynamics of our Galaxy using different tracers, \citep[e.g.,][]{Hou2014, Nakanishi2016,Xu2018a, Reid2019, Skowron2019,Cantat2020,VERA2020, Ginard2021, Nogueras2021, Poggio2021}. It is generally known that a global spiral pattern exists in the Milky Way disc \citep[][and references therein]{Shen202}. However, the finer details of the spiral arms, i.e., the number of spiral arms, their position, orientation, small-scale arm features, their thickness and formation mechanisms, are still topics of debate. Measuring precise distances is the key to resolve all problems. Recently, Gaia space-based telescope \citep{Gaia2016} has provided accurate parallax measurements for billions of stars. This database facilitates the studies of the Milky Way Galaxy in 6D phase space to explore the structure and formation of our Galaxy. \citet{Xu2018a, Xu2018b} have studied the Galactic spiral arm structure within 3 Kpc of the Sun using Gaia DR2 data \citep{GaiaDR22018} of OB stars and up to 5 kpc using Gaia EDR3 \citep{GaiaEDR32021} data \citep{Xu2021}. However, due to limitations imposed by the uncertainty in parallax measurements, it is possible to study the structure of the Galaxy/spiral arms only in the vicinity of the Sun \citep{Hou2021}. Therefore, it becomes difficult to study the complete structure/morphology of the Disc using Gaia data alone.

 Various studies have made use of different tracers to study spiral arms. The older population primarily gives an indication of outer arms \citep{Benjamin2005, Churchwell2009, Skowron2019} while the inner arms are traced by gas, dust, young stars, open clusters and star-forming regions, \citep[e.g.,][]{Hou2021,  Drimmel2022}. This discrepancy poses a question about the formation and evolution of the Milky Way disc. The age gradient across the spiral arms is also observed in external galaxies \citep[e.g.,][]{Dobbs2010,Shu2016,Shabani2018,Yu2018,Peterken2019}. In order to understand the formation mechanism completely, we should study the distribution of different age populations. The earlier results mainly focused on mapping the spiral structure using younger population \citep[][and references therein]{Hou2021}($\sim Myr$, high mass star forming regions (HMSFR), O-B stars, upper main sequence stars, and open clusters). The systematic study of a large number of older sources is required over the whole Galaxy. 
 
 Aside from the spiral features, the asymmetry in the disc structure due to warping of the disc is also observed in various tracers, \citep[e.g.,][]{Kerr1957, Henderson1979, Drimmel2001,warpHI, Lopez2002,yusifov2004,warpdust, Levine2006, Poggio2018, Chen2019, Chrobakova2022}. 
 However, the asymmetries in the spiral arms due to disc warping have not been studied so far.
 
In this paper, we present a systematic study to map the red clump stars over the whole Galactic plane, except the bulge region, to trace the spiral arms and the asymmetry arising from the warp. The paper is organized as follows. The data used in our study to extract the red clump sample are presented in Section \ref{sec:2}. Section \ref{sec:2.1} describes the automatic selection procedure to isolate red clump stars and the removal of contamination from the selected sample. Sample completeness with respect to other catalogues available in the literature and their distance comparison with our estimated values are discussed in Section \ref{sec:validate}.  The results obtained by studying the distribution of the final red clump star sample are discussed in Section \ref{sec:3}. We summarize our study with discussion and conclusions in Section \ref{sec:5}.

\section{Data used}\label{sec:2}
The Two Micron All Sky Survey \citep[2MASS,][]{Skrutskie2006} maps the entire sky in near-infrared wavebands J ($1.25 \mu m$), H ($1.65 \mu m$) and $K_s$ ($2.17 \mu m$) using similar telescopes and detectors in northern (Mt. Hopkins, Arizona, USA) and southern (CTIO, Chile) hemispheres. The $3\sigma$ limiting magnitudes in J, H and $K_s$ bands are 17.1, 16.4 and 15.3 mag, respectively, with S/N = 10. We have used best quality 2MASS point source data (with qflag = AAA) in order to select red clump stars in different lines of sight. The stars with J-magnitude brighter than 14.5 mag and photometric uncertainty less than 0.1 mag (in J, H, and $K_s$) are considered in our analysis. 

We also used data from Gaia space-based telescope. Gaia early data release (EDR3) \citep{GaiaEDR32021} has the data for the first 34 months of its operational phase. It contains information on 1.8 billion sources ranging from 3 to 21 G-band magnitude. This data release gives full astrometric data \citep{GaiaEDR3astrometry} of 1.4 billion sources. We make use of the distance information of Gaia EDR3 stars  provided in \citep{BJones2021}, and astrophysical parameters from Gaia data release 3 (DR3) \citep{gspphot} to remove foreground and evolved stars contamination.

\section{Red Clump Sample selection}\label{sec:2.1}
The red clump (RC) stars are low mass, core helium burning stars having a narrow range of intrinsic luminosity and colour at that phase \citep{Girardi2016} of their evolution. This property makes them reliable distance indicators. The luminosity function derived in \citet{Ham1999} shows that  at  the  peak  of  the  red  clump stars (K-giants), there  are  about  ten  times more sources than stars either 0.8 mag brighter or fainter. Being a good distance indicator and numerously present in the Galaxy, they can be used to probe the structure of intermediate-old Galactic disc. The RC stars are selected by analysing 2MASS colour-magnitude diagrams (CMDs) constructed from $40^\circ \le \ell \le 320^\circ$, avoiding the bulge region and $-10^\circ \le b \le 10^\circ$ with 1deg$^2$ binning, thus covering 5600 deg$^2$ fields. 

\subsection{Extraction of candidate RC stars from 2MASS CMDs}\label{sec:2.1.1}
We build 2MASS colour magnitude diagrams ($J-K_s$, $J$) in 1 deg$^2$ region around central $\ell$ and $b$ with boundaries of $\ell -0^\circ.5 \le \ell < \ell+0^\circ.5$ and $b-0^\circ.5 \le b < b+0^\circ.5$. The majority of the stars lie in the main sequence and giant phase in the CMDs. The most common type of giants are K-giants; therefore, the redder dense region of the CMD corresponds to RC stars. The RC stars get distributed in vertical and horizontal directions of CMD due to the effect of distance and extinction along the line of sight. Hence, instead of a clump, they appear as a band in the CMDs. The stars lying on the bluer side of the RC stars are predominantly foreground dwarf stars, while the redder regions contain M-giant or AGB-type evolved stars at higher extinction and larger distance.

\begin{figure*}[!hbt]
   \centering
   \includegraphics[scale=0.27]{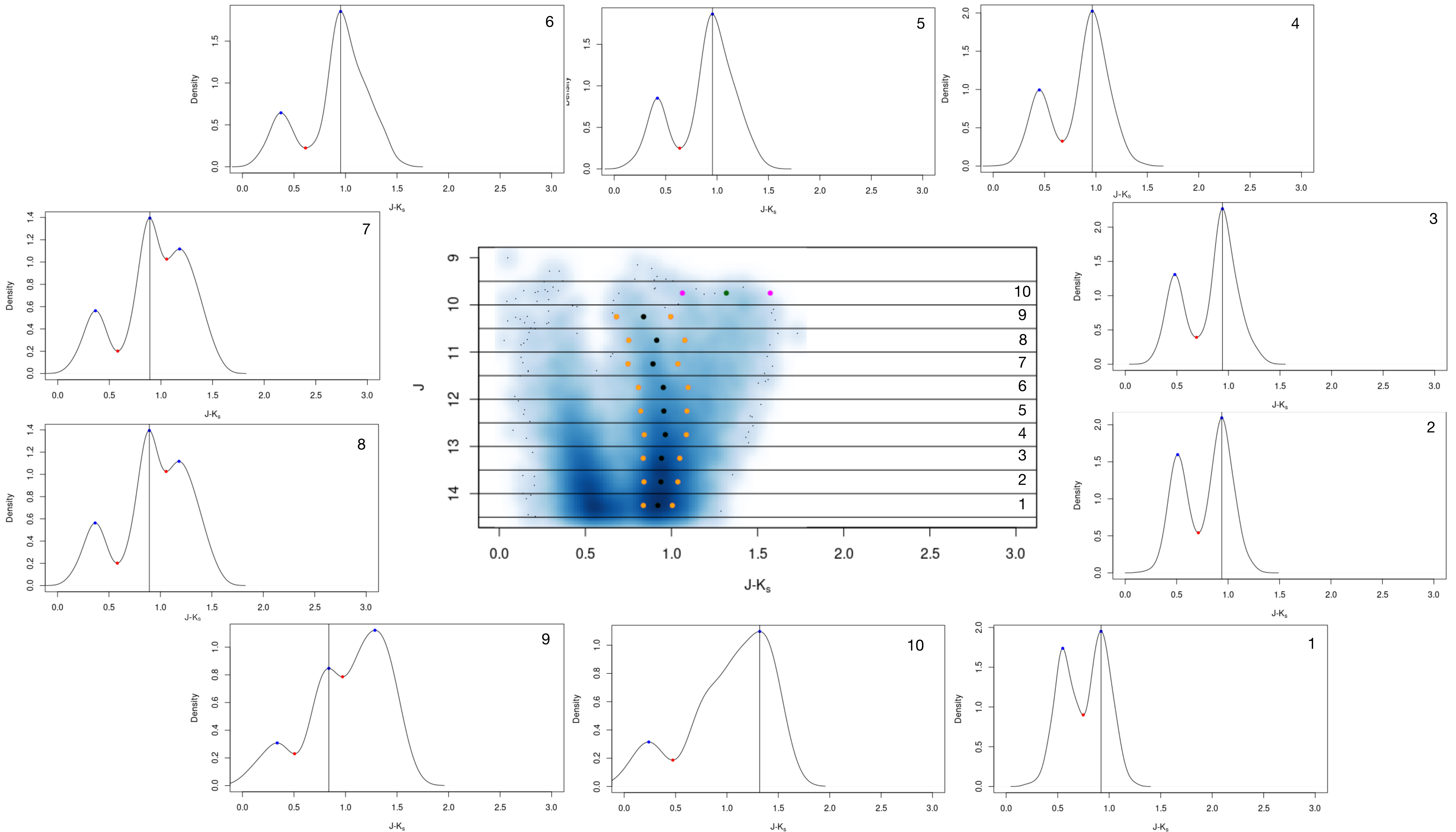}
      \caption{Selection of RC locus (black points) and its colour spread (represented in orange points) from 2MASS colour-magnitude diagram ($J-K_s, J$; central panel) for $1^\circ \times 1^\circ$ field around centred at $\ell = 45^\circ$ and $b = 5^\circ$. The figures labelled 1-10 represent $J-K_s$ density histograms of corresponding J magnitude bins.  A detailed discussion is presented in the text. The black dots in the central plot shows the last hundred points in the lowest density region.}
         \label{fig:1a}
   \end{figure*}

We made horizontal cuts in the CMD using adaptive binning in J-magnitude (see the central panel in Fig. \ref{fig:1a} for 1 deg$^2$ field centred at $l=45^\circ$ and $b=5^\circ$). The RC locus (e.g, black points in Fig. \ref{fig:1a}) is traced by selecting the  most probable values corresponding to a second peak in colour ($J-K_s$) histogram of each J-bin. Panels 1-10 of Figure \ref{fig:1a} show the density histograms of the corresponding J-bin. The $J-K_s$ colour corresponding to the second peak represents the RC locus point (black point in central panel) in that bin. The spread in the colour (orange points in the central panel of Fig. \ref{fig:1a}) around the central peak is determined from the standard deviation in the colour of all the stars present between `lo' and `hi' $J-K_s$ colour cuts in each histogram. Where `lo' and `hi' colours are calculated as 

shift = colour corresponding to second maxima - colour at first minima

lo = locus point - shift

hi = locus point + shit

\begin{figure*}[!hbt]
   \centering
   \includegraphics[scale=0.33]{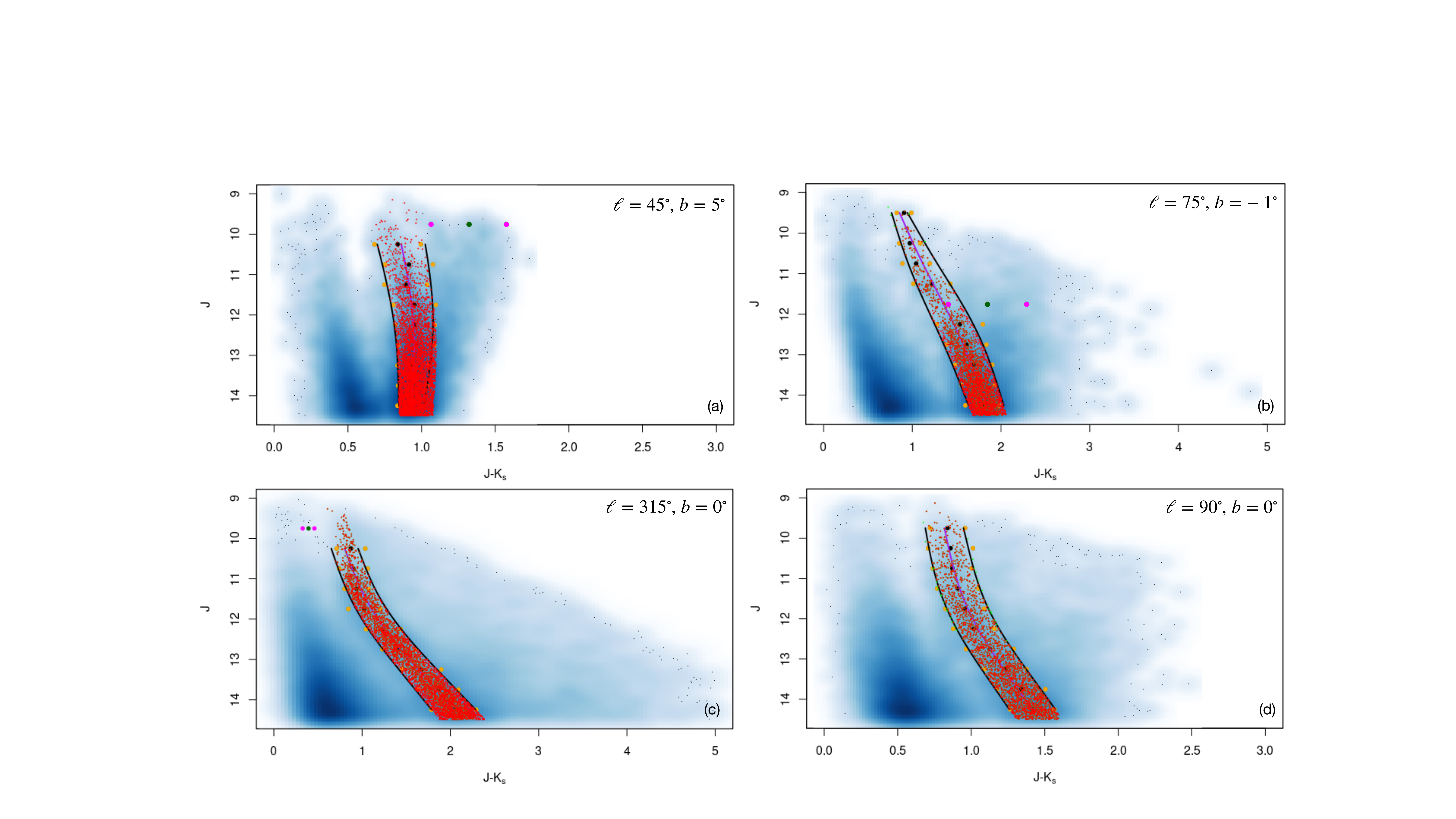}
      \caption{2MASS colour-magnitude diagram ($J-K_s, J$) of $1^\circ \times 1^\circ$ region centred at $\ell = 45^\circ, \; b= 5^\circ$; $75^\circ \; b= -1^\circ$; $315^\circ, \; b = 0^\circ$; and $90^\circ, b=0^\circ$ respectively in panel (a), (b), (c), and (d). Red points represent the selected RC star candidates. The black dots in the density plots are used to show the last 100 points in the lowest density region.}
         \label{fig:1b}
\end{figure*}
Thereafter smooth spline functions are fitted (black lines in Fig. \ref{fig:1b}) on points 1$\sigma$ region away from central RC locus (orange points in the example field) on both sides after removing the outliers (green and magenta points). All the stars within the 1$\sigma$ region are thus considered as candidate RC stars (red points in example fields shown in Fig. \ref{fig:1b}). The RC stars have an intrinsic colour spread of $\sim 0.3$ mag \citep{Indebetouw2005} due to different ages and metallicities. The chances of contamination of low-mass dwarfs increase at the faint end of the CMD. Thus, the $1\sigma$ region around the central locus is a compromise between the necessity to avoid contamination while at the same time including most of the RC stars. In certain low extinction fields (e.g. panel (a) of Fig. \ref{fig:1b}), the selection colour limits(marked by black lines) become narrower than the intrinsic colour spread in RC stars towards the fainter end.  In such cases, a colour cut of 0.15 mag is considered on the redder side of the central locus instead of the method described above. A similar cut on the bluer part is not used in order to avoid contamination from the low-mass main sequence stars. A glimpse of the RC selection in different directions is provided in  panel (a)-  $\ell = 45^\circ, \; b= 5^\circ$, (b)- $75^\circ \; b= -1^\circ$, (c)- $315^\circ, \; b = 0^\circ$ and (d)- $90^\circ, b=0^\circ$ of Fig. \ref{fig:1b}

Assuming $\frac{A_J}{A_K} = 2.5$ \citep{Lopez2002, Indebetouw2005}, the distance (d) and extinction ($A_J$) of each extracted RC star is calculated from equation \ref{eq:2} and \ref{eq:3}.
\begin{equation}\label{eq:2}
A_J = \frac{(m_J-m_{K_s})-(J - K_s)_0}{0.6} 
\end{equation}
\begin{equation}\label{eq:3}
d = 10^{\frac{m_J - M_J +5 -A_J}{5}}
\end{equation}
   
Where, $m_J$ and $m_{K_s}$ represents apparent magnitude in J and $K_s$ bands. We have assumed a Dirac delta luminosity function for K-giant stars having absolute magnitude $M_J$ and $M_{K_s}$ of $-0.945\pm 0.01$ and $-1.606 \pm 0.009$ \citep{Ruiz2018}.  The effect of metallicity on the absolute magnitude and intrinsic colours of the Sloan Digital Sky Survey's (SDSS) Apache Point Observatory Galactic Evolution Experiment (APOGEE) and the GALactic Archaeology with HERMES (GALAH) RC stars is calculated in \citet{Plevne2020} by using Bayesian-based distance from \citet{BJones2018}. The median colours of RC stars in the near IR bands $J-K_s = 0.62\pm0.05$ mag, are the same for low metallicity and high metallicity RC stars. While, the median value of absolute magnitude in the J-band shows a slight difference, i.e., $M_J = -1.05 \pm 0.21$ for low-$\alpha$ RC stars and $M_J = 0.89 \pm 0.27$ mag for high-$\alpha$ sources. The distance derived using these values of absolute magnitude and colours for the low and high metallicity in APOGEE RC sample \citep{Bovy2014} shows an offset of $\sim 200$ pc only. Hence, it is safe to assume solar neighbourhood values of absolute magnitude and intrinsic colour in near-infrared bands to calculate the distance. 

  Thus, the errors in the distance determination will include the errors in the absolute magnitude (0.5\% contribution), extinction and intrinsic colour (3\% contribution), and the errors by assuming a Dirac delta luminosity function (~5\%).  The photometric error will have a negligible contribution while using mean magnitude in a bin with more than 50 stars. Including all the uncertainties, the error in the distance determination is less than 10\%. A comparison of the distance estimated by our method with those provided by Gaia and APOGEE is discussed in section \ref{sec:Dis}.
Using this method, we have selected $\sim$ 10.5 million RC star candidates in  $40^\circ \le \ell \le 320^\circ$ and $-10^\circ \le b \le 10^\circ$ region. 

\subsection{Contamination}\label{sec:2.1.2} 
We have chosen the $1\sigma$ region around the central locus in CMD to extract RC stars. This selection reduces the contamination level, but there could be significant contamination left due to the overlapping foreground dwarf or giant stars at the fainter end of CMD. This issue is resolved by utilising the high-accuracy astrometric data from the Gaia space-based telescope. The Gaia counterparts corresponding to our selected 2MASS RC stars are obtained by matching our sample with the best neighbours of Gaia DR3 and EDR3 using \citet{crossmatch} through ADQL query. We found counterparts of 10.3 million RC stars in Gaia DR3 and EDR3 out of 10.5 million selected stars.  The RC candidate stars having fractional Gaia distance ($r_{pgeo}$) error less than 10\%  and absolute difference in $r_{pgeo}$ and distance obtained by our method more than 1.5 kpc, are assumed as foreground low mass star contamination. The resulting contamination is thus found to be $\sim 15\%$ in our sample. We also crosscheck the level of contamination using the current release of Gaia data, i.e., GAIA DR3. A total of 14\% of the sources are contaminating our sample,  i.e.,  stars having positive parallax, parallax uncertainty less than 10\% and the absolute distance difference  $|d_{our~method}- \frac{1}{parallax}| > 1.5$ kpc. This obtained contamination is removed from the sample, and the remaining stars are considered as `pure' RC stars. The final sample contains $\sim 10$ million RC stars.

A further crosscheck on the contamination and completeness in the sample is estimated by considering the latest Besan{\c{c}}on model\footnote{\href{https://model.obs-besancon.fr/}{Web service of the Besan{\c{c}}on model}} \citep{Besancon2012} around a simulated $1^\circ \times 1^\circ$ field centred at $\ell = 45^\circ$, $b=5^\circ$ using Marshall’s extinction map \citep{warpdust} in the 2MASS photometric system. The resulting model CMD has 11,539 stars down to J = 14.5 mag. Comparing the actual number of RC stars present in the field (5154) with our selected value (5133), we see that around 16\% of the RC stars are missed by our selection criteria. Also, a total of $\sim$14\% of dwarf and giant stars are contaminating our sample, which is consistent with the contamination rate determined using Gaia foreground stars.

The contamination from foreground dwarf and giant branch star in the final RC sample is further discussed in Appendix \ref{ap:2} along with its effect on overall results. 

\section{Sample validation:  Comparison with other available catalogues}\label{sec:validate}
In this section, the obtained RC sample is cross-matched with other RC star catalogues available in the literature to verify the completeness and distance determination.

\subsection{Sample completeness}
Our RC selection criteria are a balance between completeness and contamination. In order to minimize the contamination, we used the 1$\sigma$ region around the central locus. Considering a perfect Gaussian around the central locus implies that the 1$\sigma$ region will contain 66.6\% of the RC stars. Increasing the completeness limit to 2$\sigma$ or $3\sigma$ will lead to an increase in the contamination by dwarf stars as well as giants. This argument can be verified in panels 7,8 and 9 of Fig. \ref{fig:1a}, where an increase in the colour spread will result in the superposition of the second (containing RC stars) and third Gaussian (corresponding to giant population). Even if we are missing 33\% (upper bound) of RC stars, we are consistent in our selection criteria to use the 1$\sigma$ region around the central locus. Eventually, we are aiming to find the over-densities as compared to the surrounding regions of the Disc. The completeness of the sample with respect to other RC catalogues is discussed in the following subsections.
\subsubsection{APOGEE value added RC catalogue}\label{sec:APOGEE}
\begin{figure}[!hbt]
   \centering
   \includegraphics[scale=0.15]{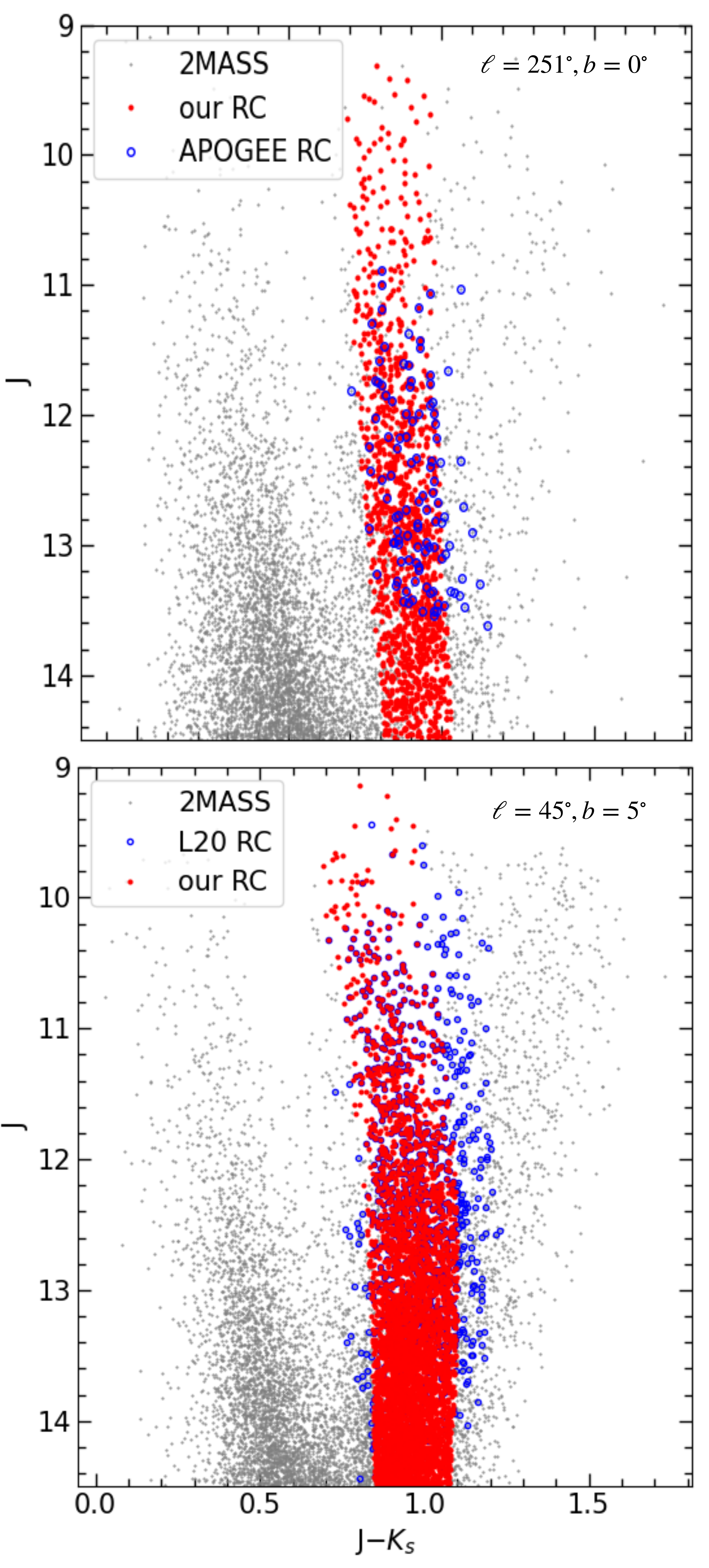}
      \caption{Colour-magnitude (J, J-$K_s$) diagram of $1^\circ \times 1^\circ$ field centred around $\ell = 251^\circ$ and $b= 0^\circ$ (upper panel) and $\ell = 45^\circ$ and $b= 5^\circ$ (lower panel) in grey-cross points. RC stars in APOGEE value-added catalogue and L20 "tier1" and "tier2" RC stars are over-plotted in blue open circles in respective panels. RC stars selected by our method are represented by red points.} 
         \label{fig:L20}
\end{figure}

The Sloan Digital Sky Survey's (SDSS) Apache Point Observatory Galactic Evolution Experiment \citep[APOGEE,][]{APOGEE} contains high-resolution near-infrared spectroscopic data of mainly  giant stars. Using this data, a highly pure (95\%) sample of RC stars is selected from the position of stars in colour-metallicity-surface gravity-temperature space by \citet{Bovy2014}.  This sample contains minimal contamination from the red giant branch (RGB), secondary red clump, and asymptotic giant branch stars. The RC value-added catalogue of APOGEE recent data release (DR17) lists 50,837 RC stars\footnote{\href{https://www.sdss.org/dr17/data_access/value-added-catalogs/?vac_id=apogee-red-clump-(rc)-catalog}{URL link to APOGEE RC catalogue}} in the entire survey region. The sample size reduces to 20,564 if we consider the stars in the region bounded by $40^\circ \le \ell \le 320^\circ$ and $-10^\circ \le b \le 10^\circ$ along with the J $<$ 14.5 mag. Out of 20,564 stars, 16,878 ($\sim$82\%) are also selected as RC candidates in our sample. Only 18\% of APOGEE RC stars are missing in our sample. In order to investigate the position of missing stars in colour-magnitude space, a J versus J-K$_s$ diagram of $1^\circ \times 1^\circ$ field around $l = 251^\circ$ and $b=0^\circ$ (containing sufficient APOGEE RC stars) is plotted in the upper panel of Fig. \ref{fig:L20}. A closer inspection of the figure reveals that the missing sources are present on the redder side of our selection window. However, increasing the selected colour range to include the missing sources may result in enhancement in the contamination.    

\subsubsection{Lucey et al., 2020 catalogue}\label{sec:L20}
Recently \citet{Lucey2020} have used available photometric data from Gaia DR2, PansTARRS, 2MASS and ALLWISE to select the RC stars. They provide a sample of 2.6 million RC stars (L20 hereafter). Their method involves predicting astroseismic ($\Delta P$, $\Delta \nu$) and spectroscopic parameters ($T_{eff}$, $logg$) obtained from SED using neural networks. They divided their RC sample into two categories:  stars with a contamination rate of 20\% and completeness of 25\% as "tier1" and 33\% contamination rate and  94\% completeness rate as "tier2". Cross-matching their catalogue with our candidate stars results in 73\% of overlapping in our survey region. In sub-classification, we are able to recover 85\% of more reliable stars from "tier1" and 71\% of tier2 stars. To have a clear idea of completeness, a colour-magnitude diagram of a low extinction field 2MASS $\ell = 45^\circ$ and $b=5^\circ$ is shown in the lower panel of Fig. \ref{fig:L20}. L20 RC stars of that region and RC candidates from our sample are over-plotted on the same figure with blue and red points, respectively. Similar to the APOGEE RC, the missing L20 RC stars are mostly present on the redder side of the RC locus. It is clear from the figure that increasing our colour selection criteria to 2$\sigma$ or 3$\sigma$ will lead to an increase in contamination.

\subsection{Estimated Distance Comparison}\label{sec:Dis}
We calculate the distance of RC stars by assuming the constant luminosity function of K-type giants within $10\%$ uncertainty level (as discussed in section \ref{sec:2.1.1}). The distance information of giant stars can also be obtained from available astrometric, spectroscopic, and photometric data. In this section, the distances obtained by different methods are compared with the estimated distance for RC stars listed in our sample.
\subsubsection{APOGEE RC and StarHorse distances}\label{sec:ApogeeDis}
\begin{figure*}[!htb]
   \centering
   \includegraphics[scale=0.4]{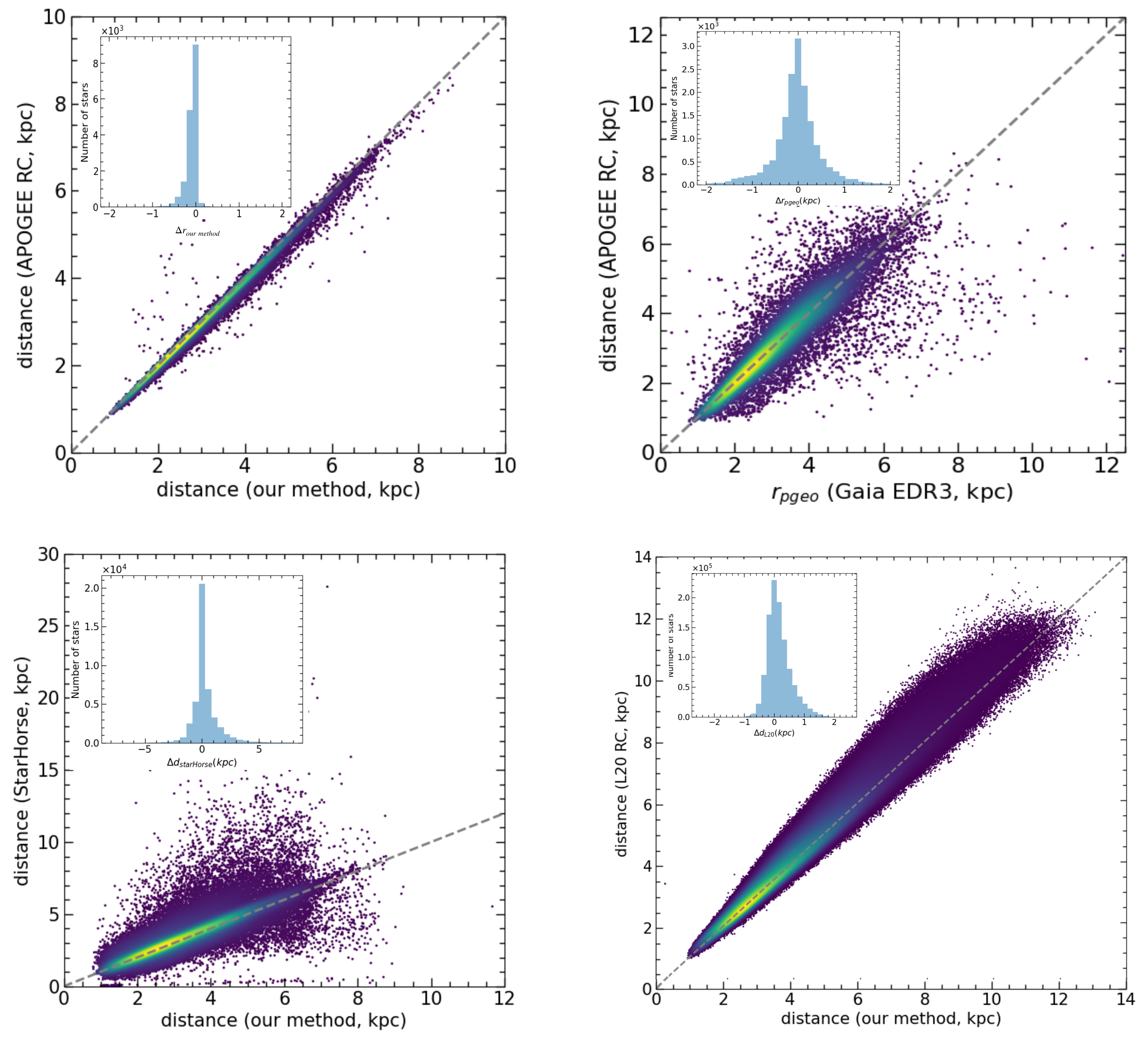}
      \caption{Comparison of distance obtain from different methods: our method with APOGEE (in the upper-left panel),  Gaia EDR3  ($r_{pgeo}$) with APOGEE RC (in the upper-right panel), our catalogue with StarHorse (bottom-left), and L20 catalogue (bottom-right). The inset histograms in each panel represent the difference in respective distances.} 
         \label{fig:dis}
\end{figure*}

The narrowness of the RC locus in colour-metallicity-luminosity space accounts for the precise measurement of the distance of stars in the APOGEE value-added RC star catalogue. The distance of common stars from the APOGEE value-added catalogue is compared with our estimated distances in the upper-left panel of Fig. \ref{fig:dis}. Most of the stars show the same distance, with a small number showing a slight deviation from APOGEE RC distances.

APOGEE has another value-added catalogue named `StarHorse' dedicated to finding the distance, extinction, and astrophysical parameters of the stars by combining the high-resolution spectroscopic data from APOGEE with broadband photometric data (Pan-STARRS, 2MASS, and ALLWISE) and parallax (Gaia EDR3) using Bayesian statistics \citep{santiago2016, starHorseDR16}. StarHorse DR17 catalogues list 562,424 (RC as well as non-RC) APOGEE sources having distance uncertainties of $\sim$ 5\%. 111,438 StarHorse stars are found in $39.5^\circ < \ell \le 320^\circ$ and $-10^\circ \le b \le 10^\circ$ region. Cross-matching these stars with our catalogue in $1^{\prime\prime}$ sky region results in 45,580 common stars. The distance comparison of cross-matched stars obtained from our method (x-axis) with StarHorse median distance is shown in the bottom-left panel of Fig. \ref{fig:dis}. The majority of the stars lie close to the 1:1 line, with some scattering of the outliers.

The distance of our selected stars matches well with both the APOGEE catalogues. However, the APOGEE spectroscopic sample has a brightness limit of H<13 mag, which corresponds to stars in nearby regions only (d $<$ 3-5 kpc). Hence, the APOGEE sample can be used to check the reliability of our catalogue distance to about $5$ kpc.
\subsubsection{Gaia Distances}
 The distance of selected RC stars can also be obtained from Gaia data, i.e., "geo" and "photogeo" parameters from \citet{BJones2021} using Gaia EDR3 data and GSP-Phot distance from Gaia DR3 \citep{gspphot}. The most appropriate distance estimator out of the three listed above was checked in \citet{Drimmel2022} (see Fig. 6 of their paper) by comparing with the RC distance from APOGEE DR17 \citep{APOGEEDR17}. They found that the "photogeo" distance gives less dispersion and hence is a better distance estimator as compared to the other two. The APOGEE RC distance of the common sample of 16,878 stars (see section \ref{sec:APOGEE}) was then compared with `photogeo/$r_{pgeo}$' in the upper-right panel of Fig. \ref{fig:dis} and distance obtained by our study in the upper-left panel along with the respective difference in distance histograms (inset Figures). It is clearly seen that our method gives a narrower distribution which is closer to the APOGEE distance as compared to the large dispersion in `$r_{pgeo}$' distance (see the inset histograms of Fig. \ref{fig:dis}). This highlights the robustness of our RC selection and distance estimation method.
\subsubsection{Lucey et al., 2020 RC distances}
APOGEE and Gaia distances can only be used to compare nearby distances (< 3-5 kpc). On the other hand, L20 catalogue provides the distances of 2.4 million stars with $\sim$ 75,000 stars at distances >10 kpc by correcting the extinction in the W1 WISE band.  A comparison of L20 distances with ours (bottom-right panel of Fig. \ref{fig:dis}) depicts that the vast majority of the stars are showing a one-to-one correlation.  The distance to the fainter stars seems to be overestimated in the L20 catalogue. This overestimate may result from an uncertain extinction in the W1 band. 

\section{Results}\label{sec:3}
In this section, we present the distribution of selected RC stars in the Galactic Disc with the aim to find over-density in spiral arms and to explore the asymmetry above and below the Galactic plane.

\subsection{Distribution in Galactic plane}\label{sec:3.1}
\begin{figure}[!h]
   \centering
   \includegraphics[scale=0.15]{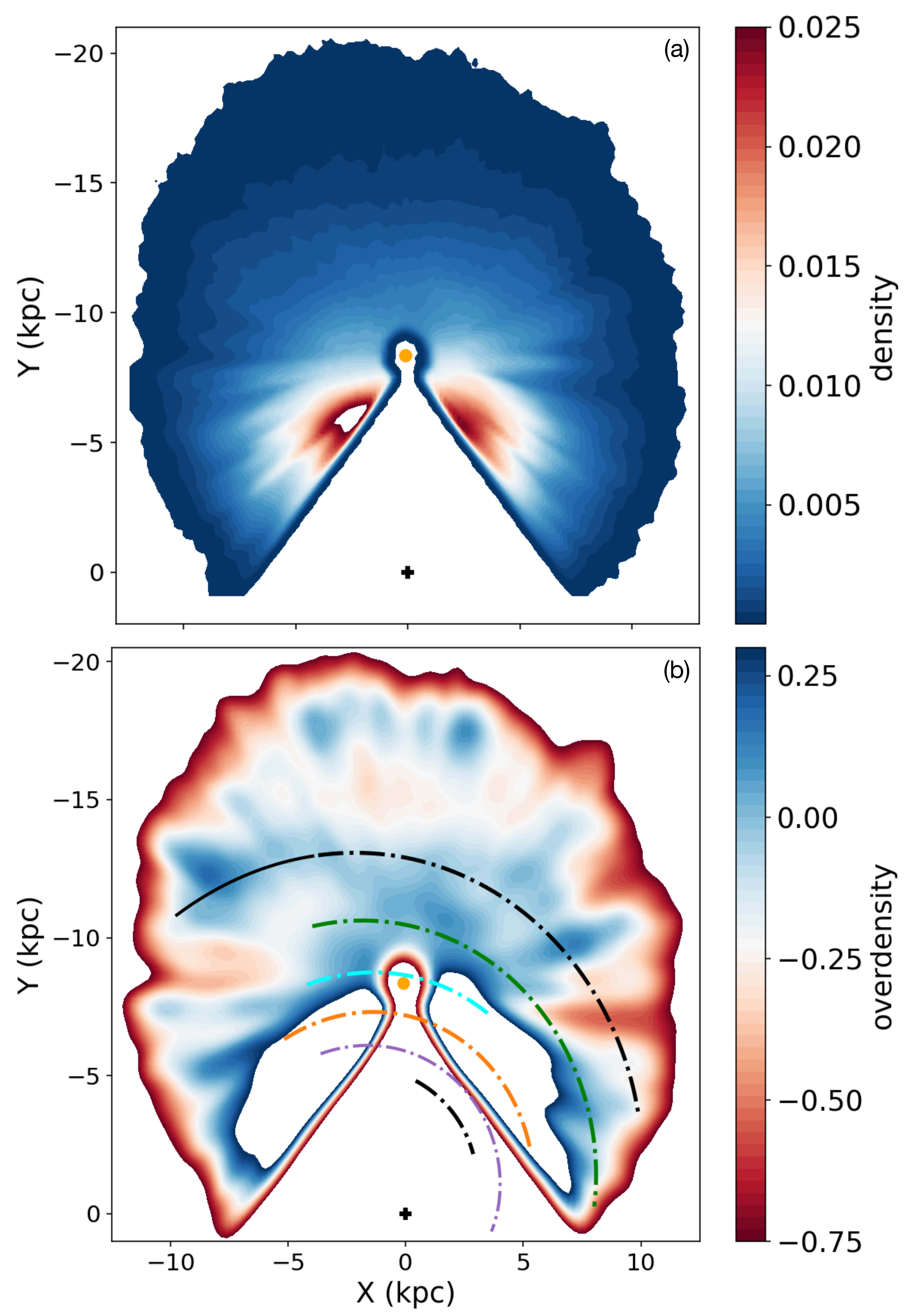}
      \caption{Spatial distribution of RC stars projected onto the Galactic plane in Galactocentric coordinates (in the top panel, (a)). (Bottom panel, (b)) Overdenity map of RC stars calculated with the local and mean density scale length of 0.3 and 2 kpc, respectively. Spiral arms from \citet{Ginard2021} (Scutum: purple, Sagittarius: orange, Local: cyan, Perseus: green) and \citet{Reid2019} (Norma-Outer: black) are overplotted in dashed-dotted lines. The solid black line corresponds to the extension of Outer arm \citep{Reid2019} in the 3rd Galactic quadrant. The empty  regions towards the inner Galaxy in both panels represent the density or over-density above the maximum limit shown in the corresponding colour bars. }
         \label{fig:2}
   \end{figure}

The complete density map of selected RC stars in the Galactocentric coordinates (X, Y) is obtained by using kernel density estimation with  0.1 kpc bandwidth.  The resulting spatial distribution of RC stars is shown in Fig. \ref{fig:2}(a), where (0,0) point represents the Galactic centre (black cross) and (-8.3, 0) corresponds to the position of the Sun (orange circle). The lack of stars closer to the Sun is due to our selection criteria discussed in previous sections.  The spatial distribution of stars does not show any arm-like feature, which is expected for older age population stars. The density distribution is rather dominated by the global density of stars in the Galactic disc i.e, exponentially decreasing star counts as moving away from the Galactic centre. The spiral arms pattern is thus difficult to disentangle from the old star population. In order to reveal the underlying spiral structure if present, we used the bi-variate kernel density estimator method. The stellar over-density is determined following \citep{Poggio2021}. 
\begin{equation}
    \Delta_\Sigma(X,Y) = \frac{\Sigma(X,Y)}{<\Sigma(X,Y)>} - 1
\end{equation}
The local density $\Sigma(X,Y)$ and mean density $<\Sigma(X,Y)>$ at (X, Y) are calculated using the bi-variate kernel density estimator using a bandwidth of 0.3 and 2 kpc, respectively. The over-density map shown in Fig. \ref{fig:2}(b) reveals some discernible arm-like features. We used the log-periodic spiral arm model from \citet{Reid2019} (eq. \ref{eq:4}) to map the spiral arms in the RC over-density plot.
\begin{equation}\label{eq:4}
    \ln\frac{R_G}{R_{G,ref}} = -(\theta_G-\theta_{G, ref})\tan \psi
\end{equation}
$R_G$ and $\theta_G$ in eq. \ref{eq:4} are Galactocentric radius and azimuth (defined as 0 towards the sun and increasing in the direction of Galactic rotation) along the arm, respectively. $R_{G, ref}$, $\theta_{G, ref}$, and $\psi$ are the reference Galactocentric radius, azimuth and pitch angle for a given arm. \citet{Ginard2021} have determined the parameters for four arms (Perseus, Local, Sagittarius and Scutum) by fitting the above model on the distribution of Galactic open cluster in Gaia EDR3 data and high mass star-forming region in the Galaxy from \citet{Reid2014parameters}. These parameters are used to construct the corresponding arm segments. The parameters of arms not considered (Norma and Outer arms) in \citet{Ginard2021} are taken from \citet{Reid2019}. The Galactocentric coordinates for corresponding arms are then calculated using the following equations.
\begin{equation}\label{eq:6}
    X = R_G \sin\theta_G 
\end{equation}
\begin{equation}\label{eq:7}
    Y = -R_G \cos\theta_G
\end{equation}
Over-plotting the spiral arm as dashed-dotted lines (Perseus: green, Local: cyan, Sagittarius: orange, and Scutum: purple, Norma-Outer: black) in Fig. \ref{fig:2} reveals that the Outer arm (black dashed-dotted line) defined in \citet{Cantat2020} is coinciding with the RC stars over-density. We note that the RC stars are tracing the Outer arm in the third quadrant (solid black line: plotted by extending the same model parameters) from (X, Y)$\sim$ (-4,-13) to (-10, -10.5) kpc as well. Most of the earlier studies of Galactic structure by star count method using different tracers either do not show such a distant Outer arm structure \citep[][etc.]{Poggio2021, Zehao2022} or show the presence of Outer arm mainly in the second quadrant with a small extension into the third quadrant (from (X, Y)$\sim$ (10, -3) to (-4, -13) kpc) \citep{Reid2019}, while  our selected RC stars are able to probe the Outer arm in farther regions of the 3rd Galactic quadrant.  The regions closer to the Perseus arm also show enhanced number density. Overall, we are able to trace the distant arms in the outer Galaxy, unlike in \citet{Zehao2022} where only the Local arm of the Galaxy was traced using RC stars extracted from the 2MASS colour-magnitude diagram. The use of Gaia parallax to estimate the distance in \cite{Zehao2022} and their selection criterion on relative parallax uncertainties $\frac{\bar{w}}{\sigma_{\bar{w}}}  > 5$ resulted in the confinement of structure towards the nearer regions of the Galaxy, thereby limiting their ability to trace structures beyond the Local arm. However, it should be noted that we have excluded the large portion of the inner Galaxy ($|\ell| < 40$), which explains why the Local arm feature is not visible in our over-density maps.
\subsection{Asymmetry in the Galactic Disc}\label{sec:3.2}
\begin{figure}[!h]
   \centering
   \includegraphics[scale=0.32]{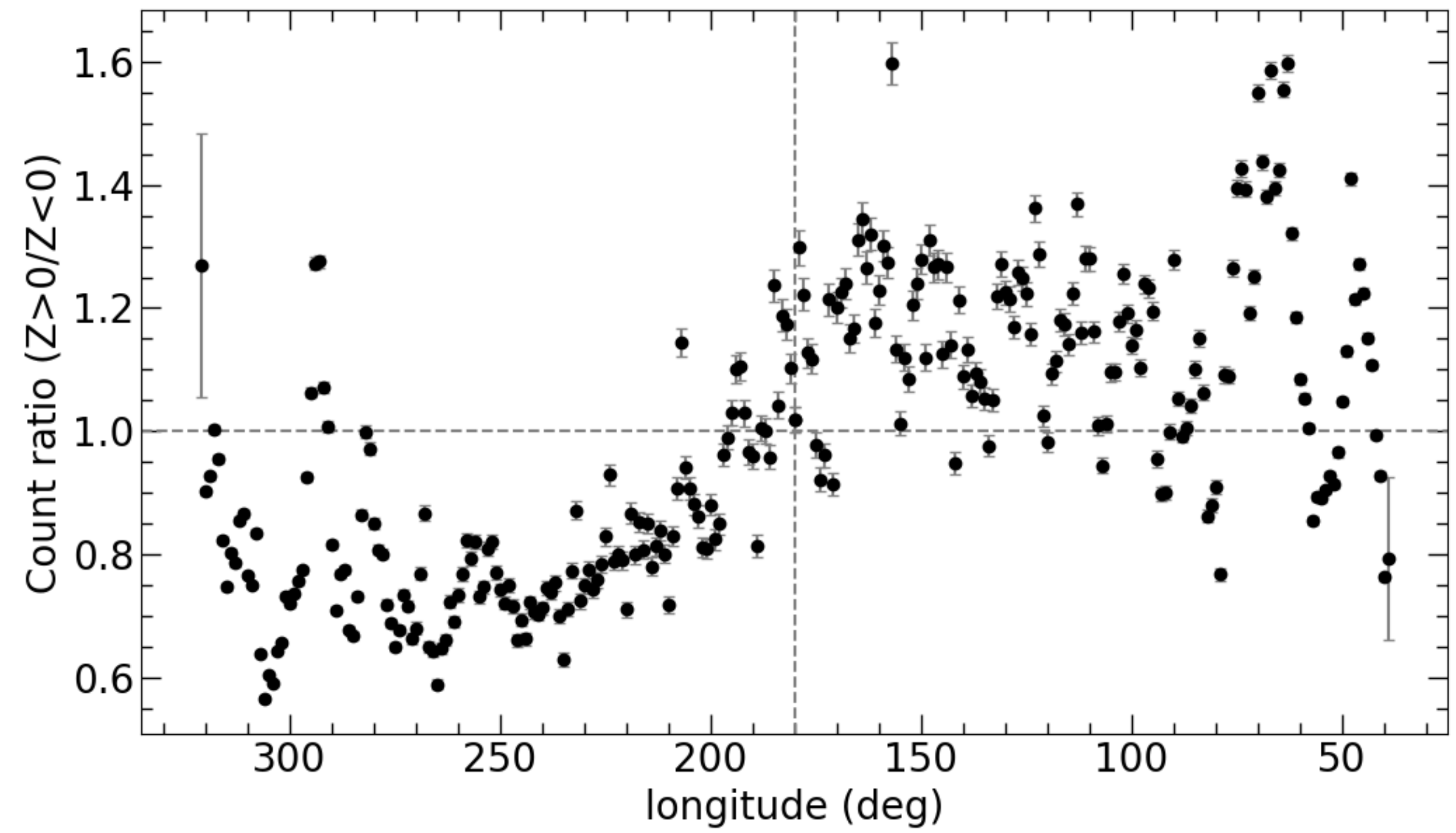}
      \caption{Distribution of the ratio of RC stars above and below the Galactic plane as a function of longitude in $1^\circ$ bins of $\ell$.}
         \label{fig:3}
   \end{figure}

The star counts above and below the Galactic plane should be nearly the same in all the longitudes for a symmetric Galaxy. However, the ratio of the number of RC stars above ($Z>0$) and below ($Z<0$) the Galactic plane in the longitude range of $40^\circ \le \ell \le 320^\circ$ with $1^\circ$ bins do not show a symmetric structure (see Fig. \ref{fig:3}). It is clearly seen that the large-scale variation of count ratio represents a wave-like asymmetry, i.e., excess of RC stars above the Galactic plane as compared to $Z<0$ for $\ell < 180^\circ$ and opposite trend for $\ell > 180^\circ$. 
The asymmetry above and below the Galactic plane with respect to longitude has also been observed earlier in a few regions of the Galactic plane using various stellar tracers, \citep[e.g,][]{Lopez2002, Ferguson2017}. We studied the continuous asymmetry, which is a manifestation of the warped disc of the Galaxy, with respect to longitude from $40^\circ \le \ell \le 320^\circ $. 

\subsection{Asymmetry in the Spiral arms} \label{sec:3.3}
 \begin{figure*}[!h]
   \centering
   \includegraphics[scale=0.25]{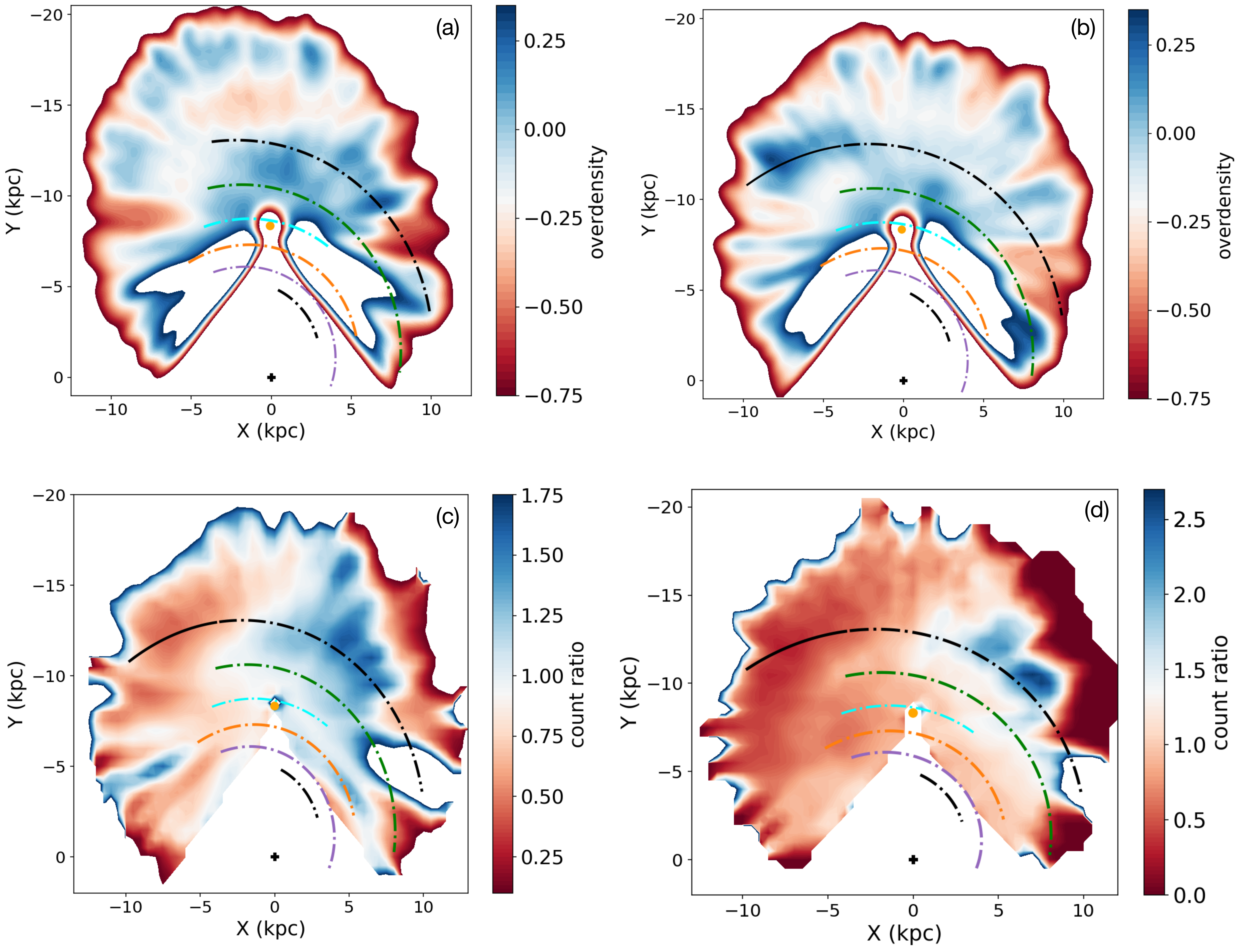}
      \caption{(Top panels) Over-density distribution of RC stars above the Galactic plane (Z>0, in panel (a)) and below the Galactic plane (Z < 0) in panel (b). (Bottom panels) Distribution of the ratio of RC stars above and below the Galactic plane in Galactocentric coordinate system with $\ell$ range of $40^\circ \le \ell \le 320^\circ$ and $-10^\circ \le b \le 10^\circ$ (in panel (c)) and $|b| = 5^\circ$ in panel (d). The spiral arm pattern (dot-dashed) follows the same colour codes as in Fig. \ref{fig:2}}
         \label{fig:5}
   \end{figure*}
It is well known that there is an asymmetry in the stellar densities along the azimuthal direction in the Milky Way as well as other disc galaxies \citep[e.g,][]{Henry2003} due to different spiral arms present in the galaxies. However, the asymmetry in the spiral arms above and below the Galactic plane has never been studied so far. The over-density of RC stars above (Z > 0, Fig. \ref{fig:5}(a)) and below the Galactic plane (Z < 0, Fig. \ref{fig:5}(b)) do not trace the Outer arm continuously from one quadrant to another.  However,  there is an over-density towards the Outer arm in the 2nd quadrant for Z > 0, and the distribution of RC stars below the Galactic plane shows over-density at the Outer arm in the third Galactic quadrant. The different parts of the Outer arm are thus traced by the RC above and below the Galactic plane, giving the signature of warping in the spiral arm, especially the outer arm. In order to see the warping in the spiral arms more clearly, we determined the ratio of RC stars above and below the Galactic plane and distributed them in the Galactocentric XY plane with 200pc bins in X and Y (see Fig. \ref{fig:5}(c) for $-10^\circ \le b \le 10^\circ$ and \ref{fig:5}(d) for $b= \pm 5^\circ$). In each panel, the large-scale wavelike asymmetry is well observed from different colours for $\ell < 180^\circ$ and $\ell > 180^\circ$.  A careful inspection of the figures reveals a curvy feature in the count ratios. The log-periodic spiral arm model described in equation \ref{eq:4} along with equation \ref{eq:6} and \ref{eq:7} is overplotted on the count ratio map to inspect the correlation of spiral arms with these curvy features. The Outer arm is coinciding with a count ratio greater than 1 in the second quadrant and a count ratio less than 1 in the third quadrant. This suggests that there is an asymmetry in the spiral pattern above and below the Galactic plane. The observed asymmetry in the spiral arms gives direct observational evidence that the spiral arms are warped upward for $\ell < 180^\circ$ and downwards in the southern direction, $\ell > 180^\circ$, similar to the Galactic disc.

\section{Discussion and Conclusion}\label{sec:5}
We developed an automated procedure to select candidate RC stars from 2MASS colour-magnitude diagrams in the Galactic disc ($40^\circ \le \ell \le 320^\circ$, and $-10^\circ \le b \le 10^\circ$) avoiding the regions closer to the bulge ($|\ell|  < 40^\circ$). A large sample of RC stars ($\sim$ 10 million) is retrieved from the 2MASS data with distance uncertainties less than 10\% and contamination rate $< 20\%$.  The spatial density distribution of RC stars in the Galactic disc is dominated by exponentially decreasing density. However, we have found an over-density in the RC stars coinciding with the Outer arm and Perseus arm. The Outer arm is traced, for the first time, continuously in the second as well as the third Galactic quadrants up to (X,Y) $\sim$ (-10, -10.5) kpc using RC stars. Significant enhancement in the stellar counts is also observed in the Perseus arm. This implies that the RC stars, representing the older age population, are good tracers to map the outer arms. This is also supported by the previous observations in \citet{Benjamin2005, Churchwell2009}. The inner arms could perhaps be traced by the younger population only. The difference in the spiral structure traced by the younger and older population could possibly represent the age gradient across the spiral arm, which favours the density wave theory for the spiral arm formation \citep{vallee2022}.  

Apart from the spiral structure, the distribution of RC stars above and below the Galactic plane shows a north-south asymmetry indicating a warp structure. Warping of the disc is also observed in dust \citep{warpdust}, gas \citep{warpHI}, in stars \citep{warpGaiaDR2}, OB stars \citep{warpOB}, Cepheids \citep{Chen2019} and also in older population \citep{Lopez2002, warp2massGaint,Chrobakova2022}, favouring the gravitational origin of the Disc warp.  Earlier studies \citep{Lopez2002} of RC stars showed the presence of warp towards some selected directions
only, but we present a systematic study over the whole Galactic plane showing the asymmetry in the Galactic disc due to the presence of warp. Not only do we see the number of stars increasing above or below the Galactic plane  in $\ell < 180^\circ$ and $\ell > 180^\circ$ respectively, but also we observe the asymmetry of stellar counts in spiral arms. This provides the first observational evidence that the spiral arms are also warped in a similar way as the rest of the Disc. 

\begin{acknowledgements}
    We would also like to thank the anonymous referee for the constructive comments and suggestions that improved the quality of the manuscript.
     This work presents the results from 2 Micron All Sky Survey (2MASS).  2MASS is a joint project of the University of Massachusetts and the Infrared Processing and Analysis centre/California Institute of Technology, funded by the National Aeronautics and Space Administration and the National Science Foundation. This publication makes use of data products from the European Space Agency (ESA) mission Gaia \href{https://www.cosmos.esa.int/gaia}{(https://www.cosmos.esa.int/gaia)}, processed by the Gaia
    Data Processing and Analysis Consortium (DPAC; \href{https://www.cosmos.esa.int/web/gaia/dpac/consortium}{https://www.cosmos.esa.int/web/gaia/dpac/consortium}). Funding for DPAC has been provided by national institutions, in particular, the institutions participating in the Gaia Multilateral Agreement.  
\end{acknowledgements}

%
%
\bibliographystyle{aa} 
\bibliography{ref.bib}

\begin{appendix}

\section{Contamination removal using Gaia DR3 astrophysical parameters}\label{ap:2}
Gaia DR3 provides the astrophysical parameters (effective temperature, surface gravity, metallicity, absolute $M_G$ magnitude, radius, distance, and extinction) for a homogeneous sample of 471 million sources with $G<19$ mag \citep{gspphot}. It is a part of the results from the astrophysical parameters inference system (Apsis) within the Gaia Data Processing and Analysis Consortium. The parameters were calculated by the simultaneous fitting of BP/RP spectra, parallax, and G-mag using Bayesian forward modelling based on isochrone models. One of the caveats of this approach is to disentangle/break effective temperature-extinction degeneracy for red stars as seen by Fig. 6a of \citet{gspphot}. However, high-quality parallax data, i.e., $\frac{\Bar{w}}{\sigma_{\bar{w}}} > 20$ is expected to give reliable $T_{eff}$ and $logg$ estimates for red giants or red clump stars. Using only high-quality parallax data corresponding to our RC sample stars, we found only $\sim 2\%$ contamination using $logg >4$ dex. 

\begin{figure}[!h]
   \centering
   \includegraphics[scale=0.16]{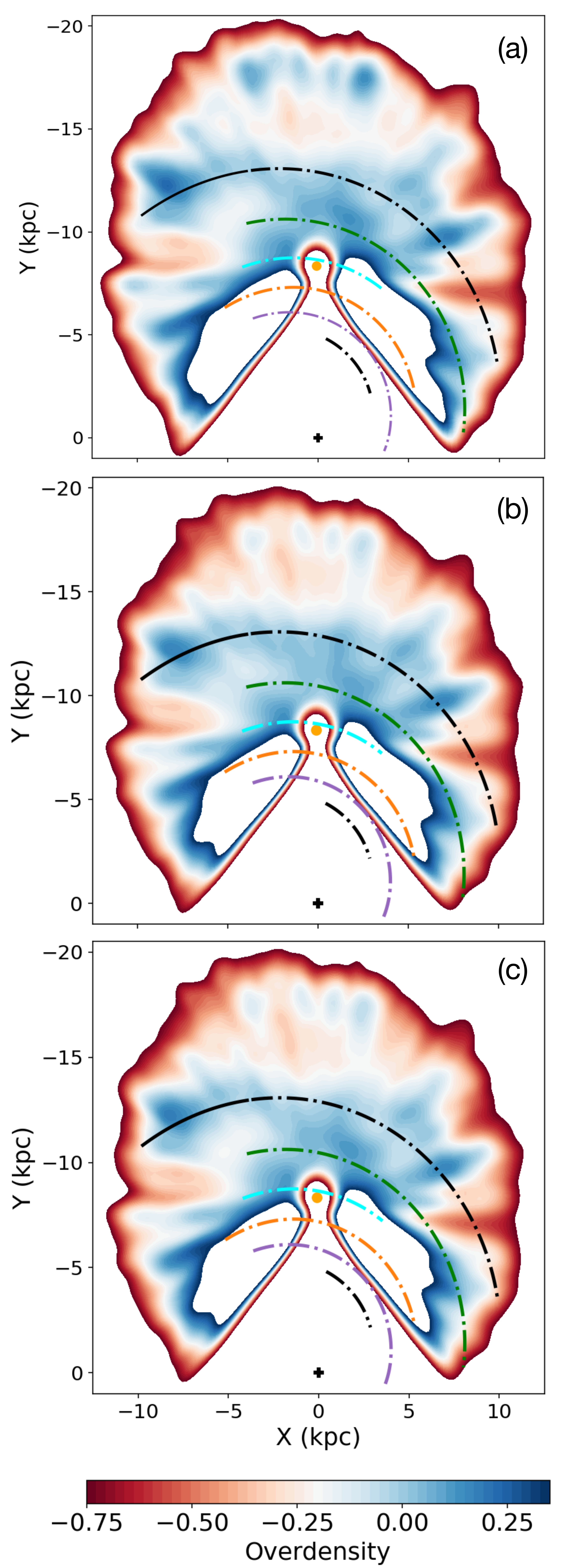}
      \caption{Comparison of RC over-density obtained by removing contamination using (a) Gaia EDR3 $r_{pgeo}$ distance and (b) astrophysical parameters from DR3. The bottom panel shows over-density by taking contributions from both (a) and (b). }
         \label{fig:A1}
   \end{figure}

For completeness, we removed the foreground contamination in our RC stars sample using $logg> 4$ dex and $T_{eff}< 6000~K$ for all stars.  This cut results in $\sim 5\%$ of contamination. The distribution of RC stars in XY plane after removing the contamination using Gaia EDR3 distance ($\sim 15\%$), Gaia DR3 astrophysical parameters ($\sim 11\%$), and from both ($\sim 20\%$) are compared in panel (a), (b) and (c) of Fig. \ref{fig:A1} respectively. The large-scale distribution of RC stars in the three panels shows similar structural features. 
\end{appendix}
\end{document}